\documentclass[conference]{IEEEtran}

\usepackage{jn_preamble_pdflatex}
\usepackage{jn_notation}
\usepackage{jn_tikz}
\usepackage{jn_tikz_coding_polar}
\usepackage{jn_tikz_coding_polarfrozenbits}
\usepackage{jn_tikz_coding_polartreechannels}
\usepackage{jn_tikz_signals_flowgraphs}
\usepackage{jn_pgfplots}
\usepackage{jn_pgfplots_coding_errorrates}
\usepackage{jn_pgfplots_coding_polarization}
\usepackage{jn_colors_parula}

\usepackage{cite}

\usepackage{refstyle}
\newcommand{\qZero}{\mathtt{0}}
\newcommand{\qpI}{\mathtt{+1}}
\newcommand{\qmI}{\mathtt{-1}}
\newcommand{\qpmI}{\mathtt{+-1}}

\newcommand{\q}{{\tn{(q)}}}
\newcommand{\unq}{{\tn{(unq)}}}

\usepackage{pifont}

\definecolor{OWNblue}{HTML}{0065BC}
\definecolor{OWNyellow}{HTML}{C9C000}
\definecolor{OWNorange}{HTML}{D37600}
\definecolor{OWNred}{HTML}{BC0001}
\definecolor{OWNpurple}{HTML}{6800D3}

\begin{document}

    \title{Ternary Quantized Polar Code Decoders:\\Analysis and Design}

    \author{%
        \IEEEauthorblockN{Joachim Neu}%
        \IEEEauthorblockA{Stanford University\\
        Email: jneu@stanford.edu}%
        \and%
        \IEEEauthorblockN{Mustafa Cemil Coşkun}%
        \IEEEauthorblockA{German Aerospace Center (DLR)\\
        Technical University of Munich (TUM)\\
        Email: mustafa.coskun@dlr.de}%
        \and%
        \IEEEauthorblockN{Gianluigi Liva}%
        \IEEEauthorblockA{German Aerospace Center (DLR)\\
        Email: gianluigi.liva@dlr.de}%
    }

    \maketitle

    \begin{abstract}
        The performance of short polar codes under successive cancellation (SC) and SC list (SCL) decoding is analyzed for the case where the decoder messages are coarsely quantized.
        This setting is of particular interest for applications requiring low-complexity energy-efficient transceivers (\eg,
        internet-of-things
        or wireless sensor networks).
        We focus on the extreme case where the decoder messages are quantized with $3$ levels.
        We show how under SCL decoding quantized log-likelihood ratios lead to a large inaccuracy in the calculation of path metrics, resulting in considerable performance losses with respect to an unquantized SCL decoder.
        We then introduce two novel techniques which improve the performance of SCL decoding with coarse quantization.
        The first technique consists of a modification of the final decision step of SCL decoding, where the selected codeword is the one maximizing the maximum-likelihood decoding metric within the final list.
        The second technique relies on statistical knowledge about the reliability of the bit estimates, obtained through a suitably modified density evolution analysis, to improve the list construction phase, yielding a higher probability of having the transmitted codeword in the list.
        The effectiveness of the two techniques is demonstrated through simulations.
    \end{abstract}

    \section{Introduction}
    \label{sec:introduction}

    Following the inclusion of polar codes \cite{phd_Stolte2002,arxiv0807.3917_Arikan2008} into the $5$th generation cellular communications standard \cite{3gpp}, the implementation of efficient polar code decoders has been gaining traction
    (\eg, \cite{ieee07527192_GiardSarkisBalatsoukasStimmingFanTsuiBurgThibeaultGross2016,arxiv1708.09603_GiardETALL2017}).
    The complexity of a decoder (and hence its power consumption and cost) can be
    reduced by coarsely quantizing
    the numeric values stored and processed within the decoder.
    Simultaneously, however, coarse quantization may deteriorate the decoder's error-correction capability.
    
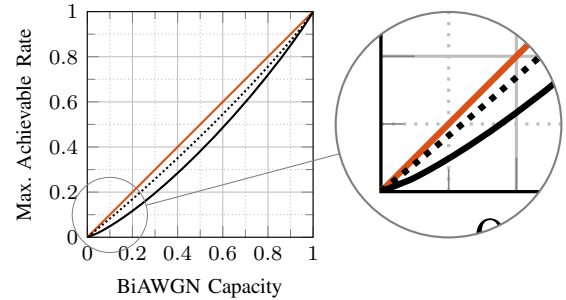
\begin{figure}%
        \centering
        \begin{tikzpicture}[spy using outlines={circle,gray,magnification=3,connect spies}]
            \begin{axis}[
                xmin=0, xmax=1,
                ymin=0, ymax=1,
                width=0.6\linewidth,
                xlabel={BiAWGN Capacity},
                ylabel={Max. Achievable Rate},
                unit vector ratio=1 1,
                legend cell align=left,
                legend style={at={(1.1,0.0)},anchor=south west,draw=none,/tikz/every even column/.append style={column sep=0.5cm},font=\footnotesize},
                grid=both,
                minor tick num=1,
                major grid style={solid,draw=gray!50},
                minor grid style={densely dotted,draw=gray!50},
                no markers,
                label style={font=\footnotesize},
                tick label style={font=\footnotesize},
            ]

                \addplot [myparula21,thick] table [x=C_BiAWGN,y=C_BiAWGN] {figures/intro-problem-statement-capacities.dat};
                \label{leg:intro-problem-statement-own-mi-biawgn}
                \addlegendentry{unq. SC: unq. $\chBiAWGN{\sigma^2}$};

                \addplot [black,densely dotted,thick] table [x=C_BiAWGN,y=C_Q3LBiAWGN] {figures/intro-problem-statement-capacities.dat};
                \label{leg:intro-problem-statement-own-mi-q3-biawgn}
                \addlegendentry{unq SC: $3$-lvl. $\chBiAWGN{\sigma^2}$};

                \addplot [black,thick] table [x=C_BiAWGN,y=R_3L_AVG] {figures/intro-problem-statement-q3.dat};
                \label{leg:intro-problem-statement-own-mi-q3-sc-dec}
                \addlegendentry{$3$-lvl. SC: $3$-lvl. $\chBiAWGN{\sigma^2}$};

                \coordinate (kink) at (axis cs:0.1,0.1);
                \coordinate (spypos) at (axis cs:1.6,0.5);

                \begin{scope}
                    \spy[size=3.0cm] on (kink) in node [fill=white] at (spypos);
                \end{scope}

                \legend{}

            \end{axis}
        \end{tikzpicture}
        \caption{Maximum achievable rate for polar codes vs. capacity of underlying BiAWGN channel; for unquantized SC decoding over unquantized BiAWGN channel \ref{leg:intro-problem-statement-own-mi-biawgn} and over $3$-level quantized BiAWGN (3Q-BiAWGN) channel \ref{leg:intro-problem-statement-own-mi-q3-biawgn}, and $3$-level quantized SC decoding over 3Q-BiAWGN channel \ref{leg:intro-problem-statement-own-mi-q3-sc-dec}.}
        \label{fig:intro-problem-statement-hassaniurbanke2012}
    \end{figure}

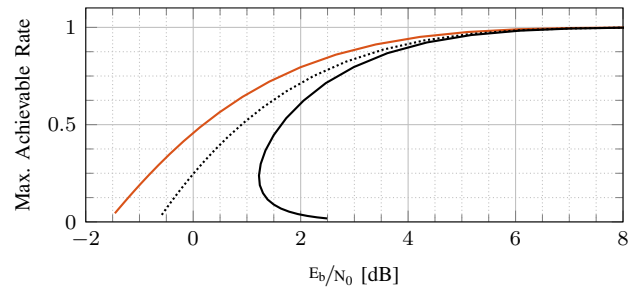
\begin{figure}%
        \centering
        \begin{tikzpicture}[spy using outlines={circle,gray,magnification=3,connect spies}]
            \begin{axis}[
                x filter/.code={\pgfmathparse{10*log10(\pgfmathresult))}},
                xmin=-2, xmax=8,
                ymin=0, ymax=1.1,
                width=0.985\linewidth,
                height=0.5\linewidth,
                xlabel={$\cEbNo$ [dB]},
                ylabel={Max. Achievable Rate},
                grid=both,
                minor tick num=3,
                major grid style={solid,draw=gray!50},
                minor grid style={densely dotted,draw=gray!50},
                no markers,
                legend cell align=left,
                legend style={at={(0.98,0.05)},anchor=south east,draw=none,/tikz/every even column/.append style={column sep=0.5cm},font=\footnotesize,nodes={text width=2cm,text depth=},cells={align=left,anchor=base},/tikz/every odd column/.style={yshift=2.5pt}},
                label style={font=\footnotesize},
                tick label style={font=\footnotesize},
            ]

                \addplot [myparula21,thick] table [x=EbN0_BiAWGN,y=C_BiAWGN] {figures/intro-problem-statement-capacities.dat};
                \label{leg:intro-problem-statement-own-mi-biawgn}
                \addlegendentry{unq. SC: unq. $\chBiAWGN{\sigma^2}$};

                \addplot [black,densely dotted,thick] table [x=EbN0_Q3LBiAWGN,y=C_Q3LBiAWGN] {figures/intro-problem-statement-capacities.dat};
                \label{leg:intro-problem-statement-own-mi-q3-biawgn}
                \addlegendentry{unq SC: $3$-lvl. $\chBiAWGN{\sigma^2}$};

                \addplot [black,solid,thick] table [x=EbN0_3L_AVG,y=R_3L_AVG] {figures/intro-problem-statement-q3.dat};
                \label{leg:intro-problem-statement-own-mi-q3-sc-dec}
                \addlegendentry{unq SC: $3$-lvl. $\chBiAWGN{\sigma^2}$};

                \legend{}

            \end{axis}
        \end{tikzpicture}
        \caption{Maximum achievable rate for polar codes vs. $\cEbNo$ of underlying BiAWGN channel
        (same curves \ref{leg:intro-problem-statement-own-mi-biawgn}, \ref{leg:intro-problem-statement-own-mi-q3-biawgn} and \ref{leg:intro-problem-statement-own-mi-q3-sc-dec} as in \figref{intro-problem-statement-hassaniurbanke2012}).}
        \label{fig:intro-problem-statement-own-ebno}
    \end{figure}

    To the best of our knowledge, so far the only work to study specifically the effect of coarse quantization on polar decoding is \cite{arxiv1209.4612_HassaniUrbanke2012}.
    The authors have chosen $3$-level (\ie, ternary) quantization after showing that it is the most extreme viable quantization, \ie, no polarization takes place for $2$-level (\ie, binary) quantization.
    It is furthermore argued that even very coarsely quantized decoding algorithms lead to excellent performance \cite{arxiv1209.4612_HassaniUrbanke2012} since successive cancellation (SC) decoding achieves a sizeable fraction of capacity for the binary-input additive white Gaussian noise (BiAWGN) channel (\cf~\figref{intro-problem-statement-hassaniurbanke2012}).
    When \figref{intro-problem-statement-hassaniurbanke2012} is reparametrized using $\cEbNo$ (with $\cEb$ being the energy per information bit and $\cNo$ the single-sided noise power spectral density, \cf\ \figref{intro-problem-statement-own-ebno}), then a considerable loss in $\cEbNo$ occurs due to quantization in the decoder, especially at moderate to low code rates (\cf\ \ref{leg:intro-problem-statement-own-mi-q3-biawgn} vs.\ \ref{leg:intro-problem-statement-own-mi-q3-sc-dec}).
    This motivates the work on low-complexity techniques to reduce the loss.

    We analyze polar code decoding with ternary quantization for short block lengths.
    In particular, we study the impact of quantization on SC and SC list (SCL) decoding.
    We show that quantized log-likelihood ratios (LLRs) lead to quantized path metrics (PMs), both of which impair bit estimation and list management of SCL decoders.
    We devise two novel solutions, namely i) selecting the codeword from the decoder's list based on maximum-likelihood (ML) rather than PM, and ii) utilizing statistical knowledge about the reliability of bit estimates to improve list management.
    Gains of up to
    \SI{0.9}{\decibel}
    in $\cEbNo$ at frame error rate (FER) $10^{-3}$ are demonstrated over conventional quantized SCL decoding for a fixed list size.

    In \secref{preliminaries}, preliminaries are provided.
    Then, we present quantized SC and SCL decoders and introduce the first innovation in \secref{q-decoding}.
    In \secref{list-enhancement}, our second novel technique is provided.
    After providing simulation results in \secref{empirical-evaluation},
    the paper concludes with a summary and outlook.

    \section{Preliminaries}
    \label{sec:preliminaries}
    
    \subsection{Notation and Terminology}
    \label{sec:notation}
    
    Regular and bold letters are used for scalar and vector quantities, respectively, while lowercase and uppercase stand for constants or random variable (RV) realizations and their corresponding RVs, respectively.
    We use uppercase bold letters for matrix constants (confusion with vector RVs is avoided by context), while calligraphic uppercase letters denote sets.
    For example, $\lambda$ denotes an LLR value, $\Lambda$ the corresponding random variable, and $\CL$ the underlying LLR alphabet.
    Vectors are column vectors, $\Bx := \left( x_0, ..., x_{n-1} \right)^\laT$,
    $\Bx_i^j := \left( x_i, ..., x_{j-1} \right)^\laT$,
    and $\Bx^j := \Bx_0^j$.
    Probability densities are denoted as $\pp[XY|Z]$ or $\pp*[x,y|z]$, mass functions as $\pP[XY|Z]$ or $\pP*[x,y|z]$,
    the expected value as $\pE{X}$.
    By $\iC{\pP[Y|X]}$ we denote the capacity of a channel with law $\pP[Y|X]$.

    \subsection{Polar Codes}
    \label{sec:polar-codes}

    Assume $n$ instances of a binary-input symmetric memoryless (BMS) channel $\pp[Y|X]$ are provided.
    The polar transform $\BG_m$
    is defined \cite{phd_Stolte2002,arxiv0807.3917_Arikan2008} as
    \begin{IEEEeqnarray}{C}
        \BG_m := \BF^{\otimes m} \BP_m^{\tn{(bitrev)}}
        \qquad\text{with}\qquad
        \BF :=
            \begin{bsmallmatrix}
                \mathtt{1} & \mathtt{1} \\
                \mathtt{0} & \mathtt{1}
            \end{bsmallmatrix}
    \end{IEEEeqnarray}
    where $\BP_m^{\tn{(bitrev)}}$ is the bit-reversal permutation,
    and $\BF^{\otimes m}$ denotes the $m$-fold Kronecker product of $\BF$. Using $\BG_m$,
    we obtain $\pp[\BY|\BU][\By|\Bu] := \pp[\BY|\BX][\By|\BG_m\Bu]$.
    Under the assumption of independent and identically distributed (i.i.d.) uniform $U_i$,
    the $i$-th synthetic channel is defined as
    \begin{IEEEeqnarray}{rCl}
        \pp[\BY\BU^i|U_i][\By,\Bu^i|u_i] &:=&
            \sum_{\Bu_{i+1}^n\in\set{\mathtt{0},\mathtt{1}}^{n-i-1}} \frac{1}{2^{n-1}} \pp[\BY|\BU][\By|\Bu].
        \IEEEeqnarraynumspace
    \end{IEEEeqnarray}

    A polar code with block length $n := 2^m$ and dimension $k$ is designed by selecting the indices of $k$ synthetic channels into a set $\CI$.
    The $k$ most reliable synthetic channels can be identified, \eg, using density evolution \cite{arxiv0901.2207_MoriTanaka2009}.
    For encoding, the matrix $\{\BG_m\}_{\CI}$ composed of the columns of $\BG_m$ indexed by the elements of $\CI$ is used as generator matrix.%

    \subsection{Successive Cancellation Decoding}
    \label{sec:sc-decoding}

    The synthetic channels lend themselves to efficient successive decoding,
    as decoding $u_0$ requires only $\By$,
    decoding $u_1$ requires $\By$ and knowledge of $u_0$,
    decoding $u_2$ requires $(\By, u_0, u_1)$,
    and so forth.
    This yields the successive cancellation (SC) decoder.
    Upon observing channel output $\By$, the $i$-th bit $u_i$ is estimated from the sign of the corresponding LLR $\lambda_i$, defined as
    \begin{IEEEeqnarray}{C}
        \label{eq:llr-computation-sc-decoder}
        \lambda_i
        :=
        \log\left( \frac{\pp[\BY\BU^i|U_i][\By,\hat{\Bu}^i|\mathtt{0}]}{\pp[\BY\BU^i|U_i][\By,\hat{\Bu}^i|\mathtt{1}]} \right).
    \end{IEEEeqnarray}
    The computations of $\lambda_i$ and $u_i$ can be defined
    recursively \cite{arxiv0901.2207_MoriTanaka2009}
    to achieve an efficient complexity of $O(n \log(n))$.

    \subsection{Density Evolution Analysis of SC Decoding}
    \label{sec:density-evolution-analysis}

    \begin{figure}[t]
        \centering
        \begin{tikzpicture}[xscale=0.7,yscale=0.7]
            \foreach \i in {0,...,7}{
                \pgfmathtruncatemacro{\inode}{\i+1};
                \node (W\i) at (\i,0) [] {$\lambda_{\tn{ch}_{\i}}$};
            }
    
            \tikzplotpolartreechannellayer[0.0]{1}{4}{P0}{W}{polarCN}{};
            \tikzplotpolartreechannellayer[0.0]{2}{2}{P1}{P0}{polarVN}{};
            \tikzplotpolartreechannellayer[0.0]{3}{1}{P2}{P1}{polarVN}{};
    
            \coordinate (Out) at ({3.5},4) {};
            \draw[-latex] (P20) -- (Out) node[midway,right] {$\lambda_{\mathtt{011}_2} == \lambda_{3}$};
    
            \draw [decorate,decoration={brace,amplitude=5pt,mirror},xshift=0,yshift=-0.6cm] (-0.3,0) -- (7.3,0) node[midway,below=6pt] {\footnotesize Channel output LLRs (i.i.d.)};
    
            \node (L0) at (-3,1) [anchor=west] {$b_2=\mathtt{0}/\bpCN$};
            \node (L1) at (-3,2) [anchor=west] {$b_1=\mathtt{1}/\bpVN$};
            \node (L2) at (-3,3) [anchor=west] {$b_0=\mathtt{1}/\bpVN$};
    
            \draw [dotted,thin] (-0.7,1) -- (7.3,1);
            \draw [dotted,thin] (-0.7,2) -- (7.3,2);
            \draw [dotted,thin] (-0.7,3) -- (7.3,3);
    
            \draw [decorate,decoration={brace,amplitude=5pt},xshift=0,yshift=0] (-3.2,0.7) -- (-3.2,3.3) node[midway,left=6pt,align=center,anchor=south,rotate=90,xshift=-1.5em] {\hspace{3em}\footnotesize Decoding layers
            \\\footnotesize $\operatorname{bin}_3(3)=(b_2 b_1 b_0)_2=\mathtt{011}_2$};
        \end{tikzpicture}
        \caption{Decoding tree for computation of $\lambda_{\mathtt{011}_2} == \lambda_{3}$ for $m=3$.}
        \label{fig:annotated-decoding-tree}
    \end{figure}
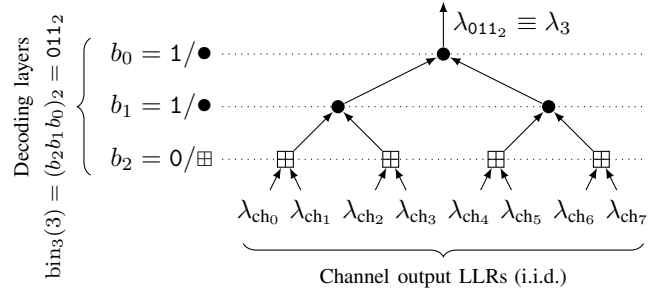

    The performance of SC decoding is analyzed using density evolution.
    The distribution of $\Lambda_i$ is obtained
    under two assumptions, namely
    i) the all-zero codeword is transmitted, and
    ii) the SC decoder is genie-aided, \ie, rather than using $\hat{\Bu}^i$ in the computation of $\lambda_i$ (\eqref{llr-computation-sc-decoder}), a genie provides $\Bu^i$.

    The recursive computation of $\lambda_i$ is then equivalent to a message-passing procedure over the decoding tree (\cf\ \figref{annotated-decoding-tree}),
    constructed as follows:
    The $2^m$ channel output LLRs $\lambda_{\tn{ch}_{i}}$ are the leaf nodes of a perfect binary tree of height $m$.
    The interior nodes are annotated with either $\bpCN$ or $\bpVN$ depending on the binary expansion $\operatorname{bin}_m(i) := (b_{m-1} ... b_0)_2\in\set{\mathtt{0},\mathtt{1}}^m$ of $i$ of length $m$, with $b_0$ being the least significant bit.
    An interior node of depth $d$, which belongs to the $(m-d)$-th decoding layer, is annotated with $\bpCN$ if $b_d=\mathtt{0}$ and with $\bpVN$ if $b_d=\mathtt{1}$.
    Each interior node applies the variable ($\bpVN$) or check node ($\bpCN$) operation on the two incoming LLR messages, where
    \begin{IEEEeqnarray}{rCl}
        x_1 \bpVN x_2   &:=&   x_1 + x_2,   \label{eq:sc-decoding-operations-perfect-vn}\label{eq:sc-decoding-operations-minsum-vn}\\
        x_1 \bpCN x_2   &:=&   2 \tanh^{-1}\left( \tanh\left( \frac{x_1}{2} \right) \tanh\left( \frac{x_2}{2} \right) \right)   \label{eq:sc-decoding-operations-perfect-cn}\\
                               &\approx&   \sign{x_1} \sign{x_2} \jnMin{\abs{x_1}, \abs{x_2}}   \label{eq:sc-decoding-operations-minsum-cn}
    \end{IEEEeqnarray}
    and passes its result upwards.
    The root node's output is $\lambda_i$.
    The so-called min-approximation (\eqref{sc-decoding-operations-minsum-cn})
    reduces computational complexity with a very limited performance loss
    \cite{DBLP:conf/icassp/LerouxTVG11}.
    Note that as $\Lambda_{\tn{ch}_i}$ are i.i.d., so are the messages output by each decoding layer.
    As a result, $\pP[\Lambda_{i}]$ is obtained via density evolution analysis \cite{arxiv0901.2207_MoriTanaka2009}.

    \subsection{Successive Cancellation List Decoding}
    \label{sec:scl-decoding}
    
    In SCL decoding \cite{ieee07055304_TalVardy2015,arxiv1401.3753_BalatsoukasStimmingPariziBurg2015}, several instances of an SC decoder are run in parallel, each for a different hypothesis on the past bit decisions.
    A vector containing the past bit decisions identifies a so-called decoding path.
    Each path is associated with an index $\ell$ and a path metric (PM) $\sclpm_\ell$.
    The SCL decoder starts out with one SC instance corresponding to the empty path, as there are no previous bit decisions.
    The empty path has $\sclpm_{\emptyset} := 0$.
    For each bit $u_i$, $0\leq i< n$, and each path $\ell$, the decoder computes $\lambda_{\ell,i}$ and produces the path's two possible offsprings $\ell_{\mathtt{0}}$ and $\ell_{\mathtt{1}}$ corresponding to $\hat{u}_i=\mathtt{0}$ and $\hat{u}_i=\mathtt{1}$, respectively.
    (If $i\not\in\CI$, only $\ell_{\mathtt{0}}$ is produced.)
    Their respective $\sclpm_{\ell_{\mathtt{0}}}$ and $\sclpm_{\ell_{\mathtt{1}}}$ are
    \begin{IEEEeqnarray}{rCl}
        \sclpm_{\ell_u} &:=& f_{\tn{PMU}}(\sclpm_\ell, \lambda_{\ell,i}, u)
        \qquad
        \forall u \in \{\mathtt{0},\mathtt{1}\} \label{eq:scl-decoding-pathmetriczero}
    \end{IEEEeqnarray}
    where the PM update function $f_{\tn{PMU}}(\sclpm, \lambda, u)$ is defined as
    \begin{equation}
        \label{eq:scl-decoding-pathmetricupdatefunction}
        f_{\tn{PMU}}(\sclpm, \lambda, u) := \sclpm + \log\left( 1+\exp\left( \left(-1\right)^{1-u} \lambda \right) \right).
    \end{equation}

    To mitigate computational complexity, only the $L$ paths with lowest PM are retained at any point, where $L$ is the list size.
    Upon completion, the collection of paths $\hat{\Bu}_{\ell}$ and their corresponding codewords $\hat{\Bc}_{\ell}$ is called the final list, denoted by $\CC_{\tn{list}}$.
    The SCL decoder ultimately decides for the path with lowest PM.
    For each element $\hat{\Bu}_{\ell}$ in $\CC_{\tn{list}}$, its $\sclpm_\ell$
    is related to its likelihood
    \cite[eq.~(13)]{arxiv1401.3753_BalatsoukasStimmingPariziBurg2015}, \ie,
    \begin{equation}
        \sclpm_\ell = - \log \left( \pPr{\BU = \hat{\Bu}_{\ell}|\BY = \By} \right).
    \end{equation}
    Hence, selecting the codeword with lowest PM from $\CC_{\tn{list}}$
    is equivalent to taking an ML decision within the list.
    This holds true for the unquantized SCL decoder.
    However, as we will see in \secref{q3-scl-decoding}, this is not necessarily true when the LLRs processed within the decoder are quantized.
    A lower bound on the block error probability of ML decoding (referred to as ML-LB) of a polar code can be estimated via Monte Carlo simulation of SCL decoding by artificially adding the transmitted codeword to $\CC_{\tn{list}}$ before taking a decision \cite{ieee07055304_TalVardy2015}.

    Since the relation between PM and likelihood of a path does not hold for quantized SCL decoders, we use different frame error rate (FER) definitions as metrics for list decoding:
    PM-FER refers to the FER of an SCL decoder which uses PM to select the codeword from its final list.
    In contrast, we write LML-FER when ML is used as selection criterion.
    Finally, List-FER is the list error rate, where a list error is declared whenever the transmitted codeword is not in the final list.

    \subsection{Three-Level Quantized BiAWGN Channels}
    \label{sec:quantization}
    
    Consider a binary-input additive white Gaussian noise (BiAWGN) channel with input alphabet $\CX = \{-1, +1\}$ and noise variance $\sigma^2$ per real-valued signal dimension.
    The signal-to-noise ratio (SNR) is $\cEsNo := \frac{1}{2\sigma^2}$, where $\cEs$ is the energy per codeword symbol.
    Furthermore, given a code of rate $R$, $\cEb := \frac{1}{R} \cEs$.
    Codeword bits are mapped to channel inputs $\mathtt{0} \mapsto +1, \mathtt{1} \mapsto -1$.
    For channel output $y$, the channel LLR of the BiAWGN is $\lambda_{\tn{ch}}^\unq := \frac{2}{\sigma^2} y$.

    In a $3$-level quantized BiAWGN (3Q-BiAWGN), the channel LLRs $\lambda_{\tn{ch}}^\unq$ are quantized to
    \begin{IEEEeqnarray}{rCl}
        \lambda_{\tn{ch}}^\q :=
            \begin{cases}
                \qmI      & \text{if } \lambda_{\tn{ch}}^\unq <= -\delta   \\
                \qZero    & \text{if } {-\delta} < \lambda_{\tn{ch}}^\unq < +\delta   \\
                \qpI      & \text{if } {+\delta} <= \lambda_{\tn{ch}}^\unq
            \end{cases}
            \IEEEeqnarraynumspace
            \label{eq:quantization-of-channel-llrs}
    \end{IEEEeqnarray}
    with reconstruction values $\qmI \mapsto -\Delta$, $\qZero \mapsto 0$, $\qpI \mapsto +\Delta$.
    The channel from $X$ to $\Lambda_{\tn{ch}}^\q$ can be seen as a binary error and erasure channel (BEEC).
    The quantization threshold $\delta$ is chosen to maximize the capacity of the BEEC.
    When $\lambda_{\tn{ch}}^\q$ is input to an unquantized decoder, we choose $\Delta$ to match the LLRs for the BEEC.
    When $\lambda_{\tn{ch}}^\q$ is input to a quantized decoder, we choose $\Delta := 1$.
    Experiments corroborate robustness of decoder performance to this choice.%
    \footnote{The choice of $\Delta$ does not affect quantized SC decoding, but it does, in principle, affect quantized SCL decoding in the PM update step (\eqref{scl-decoding-pathmetricupdatefunction}).
    EPMU-enhanced SCL decoding (devised in \secref{epmu}) is not affected.}

    \section{Quantized Polar Code Decoding}
    \label{sec:q-decoding}

    \subsection{Quantized SC Decoding}
    \label{sec:q3-sc-decoding}

    In \secref{sc-decoding}, we revisit how SC decoding is viewed as a message-passing procedure over trees (\cf\ \figref{annotated-decoding-tree}).
    This is a natural junction at which to separate the SC decoding algorithm (\ie, the sequence of operations) from the specifics of the underlying LLR algebra $(\CL, \bpCN, \bpVN, -)$, which is a set of possible LLR values $\CL$ with operations $\bpCN, \bpVN\colon \CL \times \CL \to \CL$ and $-\colon \CL \to \CL$.
    Given any LLR algebra $\CL$, an $\CL$-SC decoder and a corresponding density evolution are readily instantiated.
    This abstraction provides a framework
    to analyze `plain' quantized SC(L) decoders
    and more involved constructions in \secref{list-enhancement}.

    \begin{table}[t]
        \caption{Check Node \subref{tab:q-sc-dec-q3-cn-ops} and Variable Node \subref{tab:q-sc-dec-q3-vn-ops} Operations in $\CL_3$}
        \label{tab:q-sc-dec-q3-ops}
        \centering
        \begin{subtable}{0.45\linewidth}
            \centering
            \aboverulesep0mm
            \belowrulesep0mm
            \begin{tabular}{r|rrr}
                \toprule
                    \multicolumn{1}{c|}{$\bpCN$}  & $\qmI$   & \texttt{\ }$\qZero$ & $\qpI$   \\
                \midrule
                    $\qmI$   & $\qpI$   & $\qZero$ & $\qmI$   \\
                    $\qZero$ & $\qZero$ & $\qZero$ & $\qZero$ \\
                    $\qpI$   & $\qmI$   & $\qZero$ & $\qpI$   \\
                \bottomrule
            \end{tabular}
            \vspace{0.5em}
            \caption{Check node operation $\bpCN$}
            \label{tab:q-sc-dec-q3-cn-ops}
        \end{subtable}
        \hspace{1em}
        \begin{subtable}{0.45\linewidth}
            \centering
            \aboverulesep0mm
            \belowrulesep0mm
            \begin{tabular}{r|rrr}
                \toprule
                    \multicolumn{1}{c|}{$\bpVN$} & $\qmI$   & $\qZero$ & $\qpI$   \\
                \midrule
                    $\qmI$   & $\qmI$   & $\qmI$   & $\qZero$ \\
                    $\qZero$ & $\qmI$   & $\qZero$ & $\qpI$   \\
                    $\qpI$   & $\qZero$ & $\qpI$   & $\qpI$   \\
                \bottomrule
            \end{tabular}
            \vspace{0.5em}
            \caption{Variable node operation $\bpVN$}
            \label{tab:q-sc-dec-q3-vn-ops}
        \end{subtable}
    \end{table}

    The $3$-level quantized $\CL_3$-SC decoder uses $\CL_3 := \set{\qZero,\qpmI}$, with operations defined analogous to the min-sum
    rules
    (\eqref{sc-decoding-operations-minsum-vn,sc-decoding-operations-minsum-cn}), but clipped to $\CL_3$ (\cf\ \tabref{q-sc-dec-q3-ops}).
    We refer to the unquantized SC decoder as $\CL_\infty$-SC decoder.

    We compare $\CL_\infty$-SC decoding for BiAWGN and 3Q-Bi\-AWGN with $\CL_3$-SC decoding for 3Q-BiAWGN.
    Polar codes with $n\in\{128,256\}$ and $R=\onehalf$ were designed using density evolution \cite{arxiv0901.2207_MoriTanaka2009}.
    At FER $10^{-3}$, a loss of \SI{0.8}{\decibel} in $\cEbNo$ is caused by channel output quantization, and a further loss of \SI{1.2}{\decibel} in $\cEbNo$ is caused by quantized decoding.
    These losses are in the range predicted by previous asymptotic analysis (\cf\ \figref{intro-problem-statement-own-ebno}).

    \subsection{Quantized SCL Decoding}
    \label{sec:q3-scl-decoding}

    We extend the $\CL$-SC and SCL into the $\CL$-SCL decoder.
    In SCL decoding, the PM update is approximated \cite[eq.~(12)]{arxiv1401.3753_BalatsoukasStimmingPariziBurg2015}%
    \begin{IEEEeqnarray}{C}
        f_{\tn{PMU}}(\sclpm, \lambda, u) \approx \sclpm + \jnMax{0, (-1)^{1-u} \lambda}.   \IEEEeqnarraynumspace
    \end{IEEEeqnarray}
    This approximation is not suited for quantized decoding, \eg, for $\CL_3$ and $u=\mathtt{0}$ it maps both $\lambda\in\{\qZero,\qpI\}$ to the same PM update.
    We therefore use the refinement \cite[eq.~(3.8)]{arxiv1902.10395_Neu2019},
    \begin{IEEEeqnarray}{l}
        f_{\tn{PMU}}(\sclpm, \lambda, u) \approx \tilde{f}_{\tn{PMU}}(\sclpm, \lambda, u) := \\
        \sclpm + 
            \begin{cases}
                (-1)^{1-u}\lambda   & \text{if } (-1)^u\lambda < -2\ln(2) \\
                \frac{1}{2}(-1)^{1-u}\lambda + \ln(2)   & \text{if } \abs{\lambda} <= 2\ln(2) \\
                0   & \text{if } (-1)^u\lambda > +2\ln(2).
            \end{cases}    \label{eq:approximated-pmu}\IEEEeqnarraynumspace
    \end{IEEEeqnarray}
    We use the reconstruction values as $\lambda$ for quantized LLRs.
    As for the unquantized SCL decoder in \secref{scl-decoding},
    the conventional quantized $\CL$-SCL decoder selects the path with lowest PM from its final list $\CC_{\tn{list}}$.
    
    Note that quantized LLRs undergo severe distortion due to rounding and clipping.
    This carries over to PMs.
    PMs computed from imprecise LLRs do not preserve the order in likelihood among paths and become de-facto quantized.
    Both effects render PMs little useful for
    selecting a path from $\CC_{\tn{list}}$.

    We compare $\CL_\infty$-SCL for BiAWGN and 3Q-BiAWGN with $\CL_3$-SCL for 3Q-BiAWGN using the codes from \secref{q3-sc-decoding},
    varying $L\in\set{1,32,128}$.
    As expected, $\CL_3$-SCL decoding improves over $\CL_3$-SC decoding, \eg, \SI{0.8}{\decibel} in $\cEbNo$ at FER $10^{-3}$ for $n=256$
    and $R=\onehalf$.
    But the same gains hold for $\CL_\infty$-SCL vs.\ $\CL_\infty$-SC over 3Q-BiAWGN, so that the gap due to
    quantization in the decoder
    remains unaltered. %
    There are considerable gaps between List-FER and PM-FER of $\CL_3$-SCL, which suggests that often the transmitted codeword is contained in SCL's final list but not selected according to PM.

    \subsection{Quantized SCL Decoding with In-List ML}
    \label{sec:q3-sclml-decoding}
    
    A final list $\CC_{\tn{list}}$ is formed via the $\CL$-SCL decoding procedure as in \secref{q3-scl-decoding}.
    Within the list $\CC_{\tn{list}}$ of candidate codewords,
    the ML rule is applied to select the most likely codeword, \ie,
    \begin{equation}
        \label{eq:q-scl-ml-dec-ml-rule}
        \hat{\Bc}_{\tn{ML}} = \argmax_{\Bc\in\CC_{\tn{list}}} \pP*[\By|\Bc].
    \end{equation}

    Simulations show that
    in-list ML reliably achieves either i) the PM-FER of $\CL_\infty$-SCL, or ii) the List-FER of $\CL_3$-SCL (whichever is worse, both over 3Q-BiAWGN).
    For i), no improvement can be expected as $\CL_\infty$-SCL tightly matches the ML-LB.
    But then, ii) suggests that quantization causes the transmitted codeword to often be inadvertently removed from the list during decoding.
    This prompts the development of techniques to enhance list management in SCL in \secref{list-enhancement}.

    \section{List Enhancement Techniques}
    \label{sec:list-enhancement}

    \label{sec:epmu}

    We use statistical knowledge about the LLRs
    computed during unquantized decoding to modify the behavior of a quantized decoder in such a way that it mimics the behavior of the unquantized decoder as closely as possible, given only the instantaneous information contained in its quantized LLRs.
    In particular, we modify the quantized decoder to emulate the PM update step of an unquantized decoder.

    Let $f_{\Delta\tn{PM}}^{(i)}(\lambda_i, u_i)$ be the PM increment
    of a path $\ell_{u_i}$ that computed LLR $\lambda_i$ and decided for bit $u_i$.
    In an $\CL_3$-SCL decoder, $f_{\Delta\tn{PM}}^{(i)}$ only encounters six different input combinations, $\CL \times \{\mathtt{0}, \mathtt{1}\}$.
    Rather than computing $f_{\Delta\tn{PM}}^{(i)}$ using \eqref{approximated-pmu} and the reconstruction value associated with $\lambda_i$, a lookup table $\hat{f}_{\Delta\tn{PM}}^{(i)}$ is designed.
    To this end, assume
    $\pP[\Lambda_{i}^\unq\Lambda_{i}^\q]$
    was known.
    Then, the mean squared error between the PM updates in the unquantized and the quantized decoder is minimized with
    \begin{equation}
        \label{eq:list-enhancement-epmu-formula}
        \hat{f}_{\Delta\tn{PM}}^{(i)}\left(\lambda_{i}^\q, u_i\right) := \pE{f_{\Delta\tn{PM}}^{(i)}(\Lambda_{i}^\unq, u_i)|\Lambda_{i}^\q=\lambda_{i}^\q}.
    \end{equation}
    Hence the name expected path metric updates (EPMU).

\begin{figure}[t]
        \centering
        \begin{tikzpicture}[yscale=0.75,xscale=1.1]
            \foreach \i in {0,2,...,7}{
                \pgfmathtruncatemacro{\inode}{\i+1};
                \node (W\i) at (\i,-0.2) [] {\footnotesize $(\lambda_{\tn{ch}_{\i}}^\q, \lambda_{\tn{ch}_{\i}}^\unq)$};
            }
            \foreach \i in {1,3,...,7}{
                \pgfmathtruncatemacro{\inode}{\i+1};
                \node (W\i) at (\i,-0.2) [] {...};
            }

            \tikzplotpolartreechannellayer[-0.30]{1}{4}{Px0}{W}{polarCN}{};
            \tikzplotpolartreechannellayer[-0.30]{2}{2}{Px1}{Px0}{polarVN}{};
            \tikzplotpolartreechannellayer[-0.50]{3}{1}{Px2}{Px1}{polarVN}{};

            \tikzplotpolartreechannellayer[+0.30]{1}{4}{Py0}{W}{polarCN}{};
            \tikzplotpolartreechannellayer[+0.30]{2}{2}{Py1}{Py0}{polarVN}{};
            \tikzplotpolartreechannellayer[+0.50]{3}{1}{Py2}{Py1}{polarVN}{};

            \coordinate (OutX) at ({3.5-0.5},4) {};
            \coordinate (OutY) at ({3.5+0.5},4) {};

            \draw[-latex] (Px20) -- (OutX) node[midway,left] {$\lambda_{\mathtt{011}_2}^\q$};
            \draw[-latex] (Py20) -- (OutY) node[midway,right] {$\lambda_{\mathtt{011}_2}^\unq$};
        \end{tikzpicture}
        \caption{Coupling of a quantized and an unquantized decoder, which compute LLRs $\lambda_i^\q$ and $\lambda_i^\unq$, respectively, in parallel but based on the same channel realization (\ie, the $\Lambda_{\tn{ch}_i}^\q$ and $\Lambda_{\tn{ch}_i}^\unq$ are not independent).}
        \label{fig:list-enhancement-super-decoder-01}
    \end{figure}
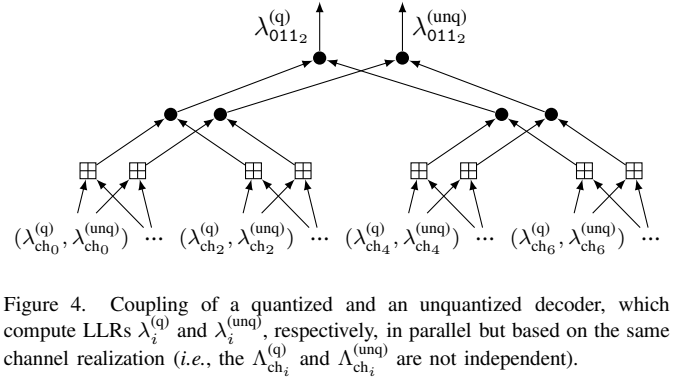

    The remainder of this section describes a construction for obtaining $\pP[\Lambda_{i}^\unq\Lambda_{i}^\q]$, based on the $\CL$-SC decoder abstraction introduced in \secref{q3-sc-decoding}.
    To this end, imagine a joint decoder composed of an unquantized and a quantized decoder (\cf\ \figref{list-enhancement-super-decoder-01}).
    Both decoders operate on the same channel realization,
    \ie, the $\Lambda_{\tn{ch}_i}^\q$ and $\Lambda_{\tn{ch}_i}^\unq$ are not independent.
    The output for the $i$-th synthetic channel
    of the joint decoder is %
    $\bigl(\Lambda_{i}^\unq,\Lambda_{i}^\q\bigr)$.
    A joint density evolution,
    under the assumptions of all-zero transmitted codeword
    and genie-aided SC decoding,
    as in \secref{density-evolution-analysis},
    is used to analyze the joint decoder in order to obtain $\pP[\Lambda_{i}^\unq\Lambda_{i}^\q]$.

    Note that the joint decoder depicted in \figref{list-enhancement-super-decoder-01} can be viewed as a $\CL_{(3,\infty)}$-SC decoder
    with $\CL_{(3,\infty)} := \CL_3 \times \CL_\infty$,
    where the LLR operations are reduced to those of the underlying quantized and unquantized decoder, \ie,
    \begin{IEEEeqnarray}{rCl}
        \bigl(x_1^\q, x_1^\unq\bigr) \bpCN \bigl(x_2^\q, x_2^\unq\bigr) &:=& \bigl(x_1^\q \bpCN x_2^\q, x_1^\unq \bpCN x_2^\unq\bigr)
            \IEEEeqnarraynumspace \\
        \bigl(x_1^\q, x_1^\unq\bigr) \bpVN \bigl(x_2^\q, x_2^\unq\bigr) &:=& \bigl(x_1^\q \bpVN x_2^\q, x_1^\unq \bpVN x_2^\unq\bigr)
            \IEEEeqnarraynumspace \\
        - \bigl(x^\q, x^\unq\bigr) &:=& \bigl(- x^\q, - x^\unq\bigr).
            \IEEEeqnarraynumspace
    \end{IEEEeqnarray}
    Furthermore, recall that under the all-zero transmitted codeword assumption, the BiAWGN channel output $Y$ is Gaussian, $Y \sim \GaussianR{1}{\sigma^2}$.
    Hence, for $\mu := \frac{2}{\sigma^2}$, $\Lambda_{\tn{ch}}^\unq \sim \GaussianR{\mu}{2\mu}$.
    When $\lambda_{\tn{ch}}^\unq$ and $\lambda_{\tn{ch}}^\q$ are computed from the same channel realization $y$, as is the case for the joint decoder at hand,
    only tuples $\bigl( \lambda_{\tn{ch}}^\q,\lambda_{\tn{ch}}^\unq \bigr) \in \CL' := \CL'_{\qmI} \cup \CL'_{\qZero} \cup \CL'_{\qpI}$
    can occur, with
    \begin{IEEEeqnarray}{rCl}
        \CL'_{\qmI} &:=& \left(-\infty, -\delta\right] \times \{\qmI\}   \\
        \CL'_{\qZero} &:=& \left(-\delta, +\delta\right) \times \{\qZero\}   \\
        \CL'_{\qpI} &:=& \left[+\delta, +\infty\right) \times \{\qpI\}.
    \end{IEEEeqnarray}
    Then,
    \begin{IEEEeqnarray}{rCl}
        \pP*[\lambda_{\tn{ch}}^\q,\lambda_{\tn{ch}}^\unq] &=& \begin{cases}
            \phi_{(\mu,2\mu)}\left(\lambda_{\tn{ch}}^\unq\right) &\text{if } \bigl(\lambda_{\tn{ch}}^\q,\lambda_{\tn{ch}}^\unq\bigr) \! \in \! \CL'   \\
            0 &\text{otherwise}
        \end{cases}
        \IEEEeqnarraynumspace
    \end{IEEEeqnarray}
    where
    $\phi_{(\mu, \sigma^2)}(x) := \frac{1}{\sqrt{2\pi\sigma^2}} \exp\left( - \frac{(x - \mu)^2}{2\sigma^2} \right)$.
    Obviously, $\Lambda_{\tn{ch}}^\q$ and $\Lambda_{\tn{ch}}^\unq$ are not independent.

    Density evolution
    is carried out
    on the $\CL_{(3,\infty)}$-SC decoder
    to obtain $\pPtilde[\Lambda_{i}^\unq\Lambda_{i}^\q]$,
    the distribution of $\BLambda_i := \bigl(\Lambda_{i}^\unq,\Lambda_{i}^\q\bigr)$
    conditional on the all-zero codeword assumption.
    The unconditional distribution is then obtained by symmetry as
    \begin{IEEEeqnarray}{rCl}
        \pP[\BLambda_{i}][\Blambda_{i}] &:=& \begin{cases}
                \frac{1}{2} \pPtilde[\BLambda_{i}][\Blambda_{i}] + \frac{1}{2} \pPtilde[\BLambda_{i}][-\Blambda_{i}] &\text{if } i\in\CI   \\
                \pPtilde[\BLambda_{i}][\Blambda_{i}] &\text{otherwise}.
            \end{cases}
        \IEEEeqnarraynumspace
    \end{IEEEeqnarray}
    EPMUs are designed from $\pP[\Lambda_{i}^\unq\Lambda_{i}^\q]$
    using \eqref{list-enhancement-epmu-formula}.%
    \footnote{Using similar joint density evolution analyses,
    extensions of EPMU can be implemented,
    \eg, the number of contradictions encountered at variable nodes
    or the number of double erasures encountered at check nodes
    can serve as a low-complexity reliability indicator to further refine EPMUs \cite[Sec.~4.2]{arxiv1902.10395_Neu2019}.}

    \section{Simulation Results}
    \label{sec:empirical-evaluation}
    
    Simulations demonstrate that the proposed techniques boost quantized polar code decoding across a wide range of scenarios.
    Our figure of merit is $\cEbNo$ at a target FER of $10^{-3}$.

    \begin{figure}[t]
        \begin{subfigure}[t]{\linewidth}
            \centering
            \begin{tikzpicture}[]
                \begin{axis}[myferplot,
                    xmode=normal, ymode=log,
                    title={Density evolution code design, $R=\nicefrac{1}{2}$, $n=256$, $L=32$},
                    xlabel={$\cEbNo$ [dB]},
                    ylabel={Frame Error Rate},
                    xmin=2.0, xmax=5.5,
                    legend columns=2,
                    height=0.65\linewidth,
                    x label style={at={(axis description cs:0.0,0.0)},anchor=south west,yshift=0em,xshift=1em},
                    y label style={at={(axis description cs:0.0,0.0)},anchor=north west,yshift=0em,xshift=0em},
                    title style={yshift=-0.5em},
                ]

                    \def\DATAPREFIX{./figures}

                    \addplotpmfer[myparula12]{\DATAPREFIX/main-01-q3biawgn-q3llrs-C_own_Q3LBiAWGN_Q3LLLR_EbN045_001_31-L32.dat}{$\CL_3$-SCL}{leg:robustness-1stlayerunquantizedthenq3-performance-pmfer-01-q3-scl-32};

                    \addplotlistfer[myparula62]{\DATAPREFIX/main-01-q3biawgn-q3llrs-C_own_Q3LBiAWGN_Q3LLLR_EbN045_001_31-L32.dat}{$\CL_3$-SCL}{leg:robustness-1stlayerunquantizedthenq3-performance-listfer-01-q3-scl-32};

                    \addplotmlinlistfer[myparula43]{\DATAPREFIX/main-03-q3biawgn-q3llrs-C_own_Q3LBiAWGN_Q3LLLR_EbN045_001_31-L32.dat}{$\CL_3$-SCL + in-list ML}{leg:robustness-1stlayerunquantizedthenq3-performance-pmfer-01-q3-sclml-32};
                    \addplotmlinlistfer[myparula44]{\DATAPREFIX/main-03-q3biawgn-epmuq3llr01s-C_own_Q3LBiAWGN_Q3LLLR_EbN045_001_31-L32.dat}{$\CL_3$-SCL + in-list ML + EPMU}{leg:robustness-1stlayerunquantizedthenq3-performance-pmfer-01-q3epmu-sclml-32};

                    \addplotpmfer[myparula22]{\DATAPREFIX/main-01-q3biawgn-unqllrs-C_own_Q3LBiAWGN_Q3LLLR_EbN045_001_31-L32.dat}{$\CL_\infty$-SCL}{leg:robustness-1stlayerunquantizedthenq3-performance-pmfer-01-unq-scl-32};
                    \addplotmllbferX[myparula21,thin]{\DATAPREFIX/main-01-q3biawgn-unqllrs-C_own_Q3LBiAWGN_Q3LLLR_EbN045_001_31-L32.dat}{$\CL_\infty$-SCL (ML-LB)}{leg:robustness-1stlayerunquantizedthenq3-performance-pmfer-01-unq-scl-32-mllb};

                    \legend{};

                \end{axis}
            \end{tikzpicture}
            \caption{FER vs. $\cEbNo$ for variants of SCL decoding over the 3Q-BiAWGN channel, demonstrating gains for in-list ML and EPMU.}
            \label{fig:robustness-1stlayerunquantizedthenq3-performance-pmfer-01}
        \end{subfigure}
        \smallskip\\
        \begin{subfigure}[t]{\linewidth}
            \centering
            \begin{tikzpicture}[]
                \begin{axis}[myferplot,
                    xmode=normal, ymode=log,
                    title={Reed-Muller code, $R=\nicefrac{37}{256}\approx0.145$, $n=256$, $L=128$},
                    xlabel={$\cEbNo$ [dB]},
                    ylabel={Frame Error Rate},
                    xmin=1.0, xmax=6.0,
                    legend columns=2,
                    xtick={1.0,1.5,2.0,2.5,3.0,3.5,4.0,4.5,5.0,5.5,6.0},
                    height=0.65\linewidth,
                    x label style={at={(axis description cs:0.0,0.0)},anchor=south west,yshift=0em,xshift=1em},
                    y label style={at={(axis description cs:0.0,0.0)},anchor=north west,yshift=0em,xshift=0em},
                    title style={yshift=-0.5em}
                ]

                    \def\DATAPREFIX{./figures}

                    \addplotpmfer[myparula12]{\DATAPREFIX/main-09-q3biawgn-q3llrs-C_rm_37_256-L128.dat}{$\CL_3$-SCL}{leg:robustness-lowrates-pmfer-01-q3-scl-128};

                    \addplotmlinlistfer[myparula43]{\DATAPREFIX/main-10-q3biawgn-q3llrs-C_rm_37_256-L128.dat}{$\CL_3$-SCL + in-list ML}{leg:robustness-lowrates-pmfer-01-q3-sclml-128};

                    \addplotmlinlistfer[myparula44]{\DATAPREFIX/main-10-q3biawgn-q3llrs-epmu-C_rm_37_256-L128.dat}{$\CL_3$-SCL + in-list ML + EPMU}{leg:robustness-lowrates-pmfer-01-q3-epmu-sclml-128};

                    \addplotpmfer[myparula22]{\DATAPREFIX/main-09-q3biawgn-unqllrs-C_rm_37_256-L128.dat}{$\CL_\infty$-SCL}{leg:robustness-lowrates-pmfer-01-unq-scl-128};
                    \addplotmllbferX[myparula21,thin]{\DATAPREFIX/main-09-q3biawgn-unqllrs-C_rm_37_256-L128.dat}{$\CL_\infty$-SCL (ML-LB)}{leg:robustness-lowrates-pmfer-01-unq-scl-128-mllb};

                    \addplotpmfer[myparula32]{\DATAPREFIX/main-09-biawgn-unqllrs-C_rm_37_256-L128.dat}{BiAWGN, $\CL_\infty$-SCL}{leg:robustness-lowrates-pmfer-01-unqunq-scl-128};
                    \addplotmllbferX[myparula31,thin]{\DATAPREFIX/main-09-biawgn-unqllrs-C_rm_37_256-L128.dat}{BiAWGN, $\CL_\infty$-SCL (ML-LB)}{leg:robustness-lowrates-pmfer-01-unqunq-scl-128-mllb};

                    \legend{};

                \end{axis}
            \end{tikzpicture}
            \caption{FER vs. $\cEbNo$ for variants of SCL decoding over the 3Q-BiAWGN channel, underlining the utility of EPMU at low code rates.}
            \label{fig:robustness-lowrates-pmfer-01}
        \end{subfigure}
        \medskip\\
        \begin{subfigure}[t]{\linewidth}
            \centering
            \footnotesize
            \tabcolsep0.25em
            \begin{tabular}{llll}
                    & \textsc{Channel} & \textsc{Decoder} & \textsc{Metric}     \\[0.2em]
                    \ref{leg:robustness-lowrates-pmfer-01-q3-scl-128} & 3Q-BiAWGN & $\CL_3$-SCL & PM-FER \\
                    \ref{leg:robustness-1stlayerunquantizedthenq3-performance-listfer-01-q3-scl-32} & 3Q-BiAWGN & $\CL_3$-SCL & List-FER \\
                    \ref{leg:robustness-lowrates-pmfer-01-q3-sclml-128} & 3Q-BiAWGN & $\CL_3$-SCL + in-list ML & LML-FER \\
                    \ref{leg:robustness-lowrates-pmfer-01-q3-epmu-sclml-128} & 3Q-BiAWGN & $\CL_3$-SCL + in-list ML + EPMU & LML-FER \\
                    \ref{leg:robustness-lowrates-pmfer-01-unq-scl-128} & 3Q-BiAWGN & $\CL_\infty$-SCL & PM-FER \\
                    \ref{leg:robustness-lowrates-pmfer-01-unq-scl-128-mllb} & 3Q-BiAWGN & $\CL_\infty$-SCL & ML-LB \\
                    \ref{leg:robustness-lowrates-pmfer-01-unqunq-scl-128} & BiAWGN & $\CL_\infty$-SCL & PM-FER \\
                    \ref{leg:robustness-lowrates-pmfer-01-unqunq-scl-128-mllb} & BiAWGN & $\CL_\infty$-SCL & ML-LB \\
            \end{tabular}
        \end{subfigure}
        \caption{FER vs. $\cEbNo$, (a) $R=\onehalf$ Polar code, and (b) $R=\nicefrac{37}{256}$ Reed-Muller code, demonstrating the gains due to the proposed techniques.}
        \label{fig:empirical-evaluation}
    \end{figure}

    \subsection{Rate \onehalf}
    \label{sec:empirical-evaluation-rate-half}

    In \figref{robustness-1stlayerunquantizedthenq3-performance-pmfer-01}, we consider the codes from \secref{q3-sc-decoding,q3-scl-decoding} with $R=\onehalf$, $n = 256$ and $L=32$.
    In-list ML $\CL_3$-SCL gains \SI{0.5}{\decibel} over conventional $\CL_3$-SCL and tightly matches its List-FER (\ref{leg:robustness-1stlayerunquantizedthenq3-performance-pmfer-01-q3-scl-32}, \ref{leg:robustness-1stlayerunquantizedthenq3-performance-listfer-01-q3-scl-32}, \ref{leg:robustness-1stlayerunquantizedthenq3-performance-pmfer-01-q3-sclml-32}).
    EPMU enables a further gain of
    \SI{0.2}{\decibel}
    by improving the List-FER (\ref{leg:robustness-1stlayerunquantizedthenq3-performance-pmfer-01-q3-sclml-32}, \ref{leg:robustness-1stlayerunquantizedthenq3-performance-pmfer-01-q3epmu-sclml-32}).
    Overall,
    \SI{0.7}{\decibel}
    of the \SI{1.2}{\decibel} losses due to quantization in the conventional decoder are reclaimed using the proposed low complexity techniques (\ref{leg:robustness-1stlayerunquantizedthenq3-performance-pmfer-01-q3-scl-32}, \ref{leg:robustness-1stlayerunquantizedthenq3-performance-pmfer-01-q3epmu-sclml-32}, \ref{leg:robustness-1stlayerunquantizedthenq3-performance-pmfer-01-unq-scl-32}).
    Finally, further gains can be achieved by increasing $L$ (\eg, for $L=128$, the gap from $\CL_3$-SCL with in-list ML and EPMU to the ML-LB of $\CL_\infty$-SCL reduces to \SI{0.2}{\decibel} \cite[Fig.~4.8]{arxiv1902.10395_Neu2019}).

    \subsection{Low Code Rate}
    \label{sec:low-code-rate}

    In \figref{robustness-lowrates-pmfer-01}, we consider the
    Reed-Muller code
    with
    $n=256$ and
    $R=\nicefrac{37}{256}\approx0.145$, which is in the low rate regime where pronounced losses due to quantization are expected (\cf\ \figref{intro-problem-statement-own-ebno}).
    At $L=128$, the losses amount to
    \SI{1.9}{\decibel}
    (\ref{leg:robustness-lowrates-pmfer-01-q3-scl-128}, \ref{leg:robustness-lowrates-pmfer-01-unq-scl-128}).
    While in-list ML alone brings no considerable gain (\ref{leg:robustness-lowrates-pmfer-01-q3-scl-128}, \ref{leg:robustness-lowrates-pmfer-01-q3-sclml-128}),
    EPMU enables a gain of
    \SI{0.9}{\decibel}
    (\ref{leg:robustness-lowrates-pmfer-01-q3-scl-128}, \ref{leg:robustness-lowrates-pmfer-01-q3-epmu-sclml-128}).

    \section{Conclusion}
    \label{sec:conclusion}
    
    We analyzed the effects of coarse quantization on SC and SCL decoding of polar codes with short block lengths.
    Quantized LLRs lead to quantized PMs, both of which impair bit estimation and list management.
    We demonstrated that in-list ML and EPMU
    can overcome these impairments, providing gains of up to
    \SI{0.9}{\decibel}
    in $\cEbNo$ at FER $10^{-3}$ over conventional quantized SCL decoding for the provided examples.

    \section*{Acknowledgements}
    \label{sec:acknowledgements}

    The authors thank Hamed Hassani, Gerhard Kramer and Rüdiger Urbanke for helpful suggestions and fruitful discussions.
    The work of J.~Neu was conducted in part during a Summer@EPFL research fellowship (\url{https://summer.epfl.ch/}).
    The work of M. Coşkun was supported by the research grant ``Efficient Coding and Modulation for Satellite Links with Severe Delay Constraints'' funded by Munich Aerospace e.V.

    \bibliographystyle{IEEEtran}
    \bibliography{own-latex-boilerplate/IEEEabrv,own-latex-boilerplate/IEEEabrv_OWN,references}

\end{document}